\documentclass[preprint,aps]{revtex4}

\usepackage{graphicx}
\usepackage{dcolumn}

\begin{document}

\preprint{Lebed}

\title{Angular Magnetoresistance Oscillations 
in Organic Conductors}

\author{A.G. Lebed$^{1,2}$, Heon-Ick Ha$^{1}$, and M.J. Naughton$^{1}$}

 \affiliation{$^1$Department of Physics, 
University of Arizona, 1118 E. 4th Street, 
Tucson AZ 85721, USA}

\affiliation{$^2$Landau Institute for Theoretical Physics,
2 Kosygina Street, Moscow, Russia}

\date{November 15, 2004, Submitted to Phys. Rev. B,  Rapid Commun.}

\begin{abstract}
 We demonstrate that electron wave functions 
 change their dimensionality at some commensurate 
 directions of a magnetic field in conductors with
 open [quasi-one-dimensional (Q1D)] sheets of 
 Fermi surface.
 These $1D \rightarrow 2D$ dimensional crossovers 
 lead to delocalization of wave functions and are 
 responsible for angular magnetoresistance 
 oscillations.
 As an example, we show that suggested theory 
 is in qualitative and quantitative agreements with 
 the recent experimental data obtained on 
 (TMTSF)$_2$ClO$_4$ 
 conductor. 
\end{abstract}

\maketitle
  
\pagebreak
  
  It is well known that, in traditional 
  "three-dimensional" metals with closed 
  quasi-particle orbits, quantum oscillations 
  in a magnetic field occur due to Landau 
  quantization of energy levels [1].
  Meanwhile, a number of low-dimensional 
  organic metals with open [quasi-one-dimensional(Q1D)] 
  sheets of Fermi surfaces (FS) [2,3], 
  where Landau quantization is not possible, 
  demonstrate several types 
  of unconventional angular oscillations 
  [4-23] related to open 
  sheets of FS,
  
\begin{equation}
\epsilon^\pm ({\bf p}) 
= \pm v_F \ (p_x \mp p_F) 
+ 2 t_b \  f (p_y b^*) + 2 t_{\perp} \cos(p_{\perp} c^*) \  , 
 \ \ \ \ p_F v_F \gg  t_b \gg t_{\perp} \ ,
\end{equation}
where $+(-)$ stands for the right (left) sheet 
of the FS; $v_F $ and  $p_F $ are Fermi velocity 
and Fermi momentum along conducting 
${\bf x}$-axis, respectively; 
$t_b$ and $t_{\perp}$ are the overlapping integrals 
between conducting chains [2,3]; 
$\hbar \equiv 1$.

Among these low-dimensional metals with Q1D sheets 
of FS (1), are (TMTSF)$_2$X (X = PF$_6$, ClO$_4$, 
ReO$_4$, ...) [8-19], (DMET)$_2$I$_3$ [20],
$\kappa$-(ET)$_2$Cu(NCS)$_2$ [21,22], (DMET-TTF)$_2$X 
(X = AuCl$_2$, ...) [23], and some other 
conductors.
[For most Q1D conductors, one can use $f(p_yb^*) = 
\cos(p_y b^*)$ in Eq.(1), with the important exceptions of 
(TMTSF)$_2$ClO$_4$ and (TMTSF)$_2$ReO$_4$ 
compounds [2,3]].

As was first recognized in theories of field-induced 
spin-density-wave phases in (TMTSF)$_2$X materials 
[24-27], in the absence of Landau quantization, 
Bragg reflections are responsible for 
quantum magnetic many-body phenomena in metals 
with open FS (1).
Moreover, as shown in Ref.[4], at some Magic Angle 
(MA) directions of a magnetic field,
 \begin{equation}
\tan \alpha =   N \ (b^* / c^*) \ , 
 \ \ {\bf H} = (0,H \sin \alpha , H \cos \alpha) \ ,
\end{equation}
electron trajectories in a reciprocal space become
periodic.
This leads to the existence of constructive 
interference effects [28,29] coming from Bragg 
reflections, which occur when electrons move 
in a magnetic field along open Q1D sheets of 
FS (1) in the extended 
Brillouin zone.
As a result, many-body angular oscillations 
in a magnetic field appear [4,5].
For extensions of the idea [4] to some model 
one-body phenomena, 
see Refs.[6,7].

 Very recently, it has been theoretically demonstrated [28-31] 
 that the similar interference effects are able to account  
 for nontrivial one-body phenomena - angular magnetoresistance 
 oscillations - experimentally observed in resistivity component 
 $\rho_{\perp}({\bf H})$, perpendicular to conducting
 planes in (TMTSF)$_2$X [8-19] and 
 $\kappa$-(ET)$_2$Cu(NCS)$_2$ [21,22]
 conductors.
 Note that the existence of angular magnetic oscillations 
 [4-23] is an a contradiction with common belief that nothing 
 important happens with open quasi-particles orbits (1) in 
 a magnetic field and, as shown in Ref.[32], is a 
 consequence of some non-trivial "momentum 
 quantization laws".
 
 In particular, in Refs.[28,29], a theory of angular 
 magnetic oscillations in $\rho_{\perp}({\bf H})$ at MA 
 directions of a magnetic field (2) was suggested, 
 whereas, in Refs.[30,31], theoretical descriptions of 
 the so-called interference commensurate (IC) oscillations 
 in $\rho_{\perp}({\bf H})$ [19,20,33-35] 
were proposed. 
The IC magnetoresistance oscillations were also studied 
by numerical methods [19,33,34].
As shown in Refs.[29,28], the physical meaning of MA 
oscillations (2) in $\rho_{\perp}(H, \alpha)$ is related 
to the interference effects between velocity component 
perpendicular to the conducting 
$({\bf x},{\bf y})$-planes, $v_{\perp}(p_{\perp}) = 
- 2 t_{\perp} c^* \sin(p_{\perp} c^*)$, 
and the density of states.
The IC oscillations are characterized by strong enough 
projection of a magnetic field on the conducting 
${\bf x}$-axis [19,20,33,34]:
 \begin{equation}
{\bf H} = H ( \cos \theta \cos \phi, + \cos \theta \sin \phi , 
\sin \theta) \ .
 \end{equation} 
 This changes the physical meaning of the interference effects 
(see Ref.[30]), although minima in $\rho_{\perp}(H, \theta,\phi)$ 
appear exactly at MA projection [19,30,31] of the field (3) 
on $({\bf y},{\bf z})$-plane, corresponding to the following
commensurate angles [30,31]:
 \begin{equation}
\sin \phi = n (b^* / c^*) \tan \theta  \ ,  
 \end{equation}
where $n$ is an integer.
 
Indeed, as shown in Ref.[30], interference effects, responsible
for the IC oscillations, occur even for constant value of density 
of states [i.e., for $v_F = const $ in Eq.(1)] and correspond
to summation of infinite number of electron waves coming from
some "stationary phase points" [30] on electron trajectories 
in the extended Brillouin zone.
Note that all existing theories of IC oscillations are developed 
for simplest Q1D spectrum with $f(p_y b^*) = \sin (p_y b^*)$ 
in Eq.(1) and, thus, are not applicable to such typical Q1D 
conductor as (TMTSF)$_2$ClO$_4$, 
where
 \begin{equation}
 f(p_y b^*)=\sqrt{ \cos^2(p_y b^*) + \Delta^2/4 t^2_b} \ ,
 \end{equation}
with $\Delta$ being the so-called anion 
gap [2,3].
 
 The main goals of our paper are as follows: 
 1) to calculate wave functions of a Q1D metal (1) 
 with arbitrary function $f(z)$ in an inclined magnetic 
 field (3), 
 2) to reveal novel phenomenon - a change of space 
 dimensionality of these wave functions from $1D$ to 
 $2D$ at commensurate directions of a magnetic field 
 (4), 
 3) to calculate $\rho_{\perp}(H, \theta, \phi)$ and to show 
 that it exhibits minima at commensurate directions (4) due 
 to the above mentioned $1D \rightarrow 2D$ delocalizations 
 of wave functions, 
 4) to demonstrate that these $1D \rightarrow 2D$ 
 delocalizations manifest themselfs as saturations of 
 magnetoresistance 
 $\rho_{\perp}(H, \theta, \phi)$, 
 5) to show that the suggested theory is in good
 agreement with the very recent experimental data 
 obtained on 
 (TMTSF)$_2$ClO$_4$ [35].

Let us discuss how $1D \rightarrow 2D$ delocalization 
crossovers can result in the appearance of 
minima in $\rho_{\perp}(H, \theta, \phi)$ using 
qualitative arguments.
 For electrons localized on conducting ${\bf x}$-chains 
 [4], resistivity component 
 $\rho_{\perp}(H, \theta, \phi)$ is expected to be 
 infinite in the absence of impurities 
 (i.e., at $1/\tau = 0$) [1,7,28]. 
 At $1/\tau \neq 0$, it has to demonstrate quadratic
 non-saturated magnetoresistance,
 $\rho_{\perp}(H, \theta, \phi)  \sim H^2 \tau $,
 in accordance with general theory [1,7] in the
 case of open quasi-particles 
 orbits (1).
If, at commensurate directions of the field (4), electrons  
become delocalized, then $\rho_{\perp}(H, \theta, \phi)$ 
is expected to have similarities with resistivity of a free 
electron 
at $H=0$. 
Therefore, in this case, $\rho_{\perp}(H, \theta, \phi)$ 
has to saturate at high magnetic fields with the saturation
values being expected to be proportional 
to $1/\tau $.
Below, we demonstrate that this qualitatively different 
behavior of $\rho_{\perp}(H, \theta, \phi)$ at commensurate 
directions (4) is indeed responsible for the appearance of 
minima in $\rho_{\perp}(H, \theta, \phi) = 
1 / \sigma_{\perp}(H, \theta, \phi)$].

To develop an analytical theory, we make use of the Peierls 
substitutions method for an open electron 
spectrum [24,4,30]:
$p_x - p_F \rightarrow - i (d/dx), \ {\bf p} \rightarrow {\bf p} 
- (e/c) {\bf A}, \ y  \rightarrow i (d/dp_y)$.
We choose the following vector potential for inclined magnetic
 field (3):
 ${\bf A} = (0, \ x \sin \theta , \ x \cos \theta 
 - y \cos \theta \cos \phi )H$, 
 where Hamiltonian (1) in the vicinity of $p_x \simeq p_F$ 
 can be expressed as
\begin{eqnarray}
&&{\hat \epsilon^{+} ({\bf p})} = 
- i v_F \biggl( \frac{d}{dx} \biggl) 
+ 2 t_b \  f \biggl[ p_y b^*- \frac{\omega_b(\theta) x}{v_F} \biggl]
+2 t_{\perp} \cos \biggl[ p_{\perp} c^*
+ \frac{\omega_c(\theta , \phi) x}{v_F}
- i  \biggl[ \frac{ {\hat \omega_c (\theta , \phi)} }{v_F} \biggl] 
\biggl( \frac{d}{d p_y} \biggl) 
\biggl] \ ,
\nonumber\\
&&\omega_b(\theta) = \frac{ e H v_F b^* \sin \theta}{ c}  \ , 
\omega_c(\theta, \phi) = \frac {e H v_F c^* \cos \theta \sin \phi}{c} \ , 
{\hat \omega_c(\theta, \phi)} = \frac{e H v_F c^* \cos \theta \cos \phi}{ c} \  .
\end{eqnarray} 

It is possible to prove that, if one represents electron 
wave functions in the form
\begin{equation}
\Psi_{\epsilon} (x, p_y, p_{\perp}) = \exp (i p_F x) 
\Psi^{+}_{\epsilon} (x, p_y, p_{\perp}) 
\ ,
\end{equation}
then solutions of the Schrodinger equation for 
Hamiltonian (6) can be written as
\begin{eqnarray}
\Psi^{+}_{\epsilon} (x, p_y, p_{\perp}) = 
&&\exp \biggl( \frac{i \epsilon x}{v_F} \biggl) 
\exp \biggl(
 - \frac{2 i t_b}{v_F} \int^x_0 
 f \biggl[ p_y b^*  -  \frac{\omega_b(\theta) u}{v_F} \biggl] 
 du \biggl)
 \nonumber\\ 
 &&\times \exp \biggl( - \frac{2 i t_{\perp}}{v_F} 
\int^x_0 
 \cos \biggl[ p_{\perp} c^*
  + \frac{\omega_c(\theta,\phi) u}{v_F}  
 + a \biggl( f \biggl[ p_y - \frac{\omega_b (\theta) u}{v_F} \biggl] 
 - f[p_y] \biggl)  
 \biggl] d u 
 \biggl) \ ,
 \nonumber\\
 &&a =  (2t_b b^*/v_F) \ .
 [ {\hat \omega_c(\theta,\phi)} / \omega_b(\theta) ] \ ,
\end{eqnarray}

Below, we demonstrate that suggested in the 
paper $1D \rightarrow 2D$ dimensional crossovers 
directly follow from Eq.(9).
For this purpose, we calculate the real space 
${\bf z}$-dependence of wave functions along the 
inter-plane direction (i.e., at $z=Nc^*$, where $N$ 
is an integer plane index) by taking a Fourier transform 
of the second exponential function 
in Eq.(9):
\begin{eqnarray}
&&\Phi^{+} (x,p_y,z=Nc^*) = 
\int_0^{2 \pi} \ \frac{d \ p_{\perp}}{2 \pi}
 \exp(i p_{\perp} N c^*) \ 
\nonumber\\
&&\times \exp \biggl( 
 - \frac{2 i t_{\perp}}{v_F} \int^x_0  
 \cos \biggl[ 
 p_{\perp} c^* + \frac{\omega_c(\theta, \phi) u}{v_F}  +
a \biggl( 
f \biggl[ p_y b^* - \frac{\omega_b(\theta)u}{v_F} \biggl] 
- f[p_y b^*] \biggl)
\biggl] d u \biggl) \ .
\end{eqnarray}
After straightforward calculations, Eq. (10) can be 
expressed as
\begin{equation}
\Phi^{+}  (x,p_y,z=Nc^*) = \exp[-i \beta N] \ 
J_{N} \biggl[ \frac{2 t_{\perp}}{v_F} \sqrt{I^2_1(x,p_y)  
+ I^2_2(x,p_y)} \biggl] \ , 
\end{equation}

\begin{eqnarray}
I_{1,2} \biggl( x = \frac{2 \pi M_0 v_F}{ \omega_b(\theta)}, p_y \biggl) =
\frac{1}{2} \biggl| \sum^{M_0-1}_{M=0} 
&&\int^{\frac{2 \pi v_F}{ \omega_b(\theta)}}_ 0 \ 
\biggl( 
\exp \biggl[
\frac{\omega_c(\theta, \phi) u}{v_F} 
+ 2 \pi M \frac{\omega_c (\theta, \phi)}{ \omega_b(\theta)}
\nonumber\\
&&+ a \biggl( 
f \biggl[ p_y b^* - \frac{\omega_b (\theta) u}{v_F} \biggl]
- f[p_y b^*]
\biggl) 
\biggl] \pm c.c. \biggl)
\ du \biggl|
\ ,
\end{eqnarray}
with $J_N (...)$ being the Bessel function [36]; $M_0$ 
is an integer, $\beta$ is a phase factor.
According to the theory of Bessel functions [36], 
$J_N(Z)$ is an oscillatory function of variable $N$ at 
$N < |Z|$, whereas it decays exponentially with $N$ 
at $N > |Z|$. 
Thus, one can conclude that wave functions (11)-(13) 
are extended along ${\bf z}$-direction if at least one 
of the functions $I_{i} (...) (i=1,2)$ in Eqs.(12),(13) is 
not restricted [i.e., if $| I_i(M_0,p_y) | \rightarrow \infty$ 
as $ M_0 \rightarrow \infty $].
In the opposite case, where both functions $I_i(...)$ 
($i=1,2$) are restricted by conditions $ | I_i (M_0
,p_y) |  < I_{max}$, 
wave functions (11)-(13) exponentially decay 
with the variable $z$ at 
$| z=N c^* | \geq 4 t_{\perp} c^* I_{max} / v_F$.

Note that functions (11)-(13) are written in the form 
of summations of infinite number of electron waves 
corresponding to quasi-classical electron 
motion in different Brillouin zones. 
Therefore, the physical meaning of summations in 
Eqs.(12),(13) is related to the interference effects due 
to Bragg reflections, which occur when electrons move 
along open orbits (1) in a magnetic field (3) in the 
extended Brillouin zone.
 As it is seen from Eqs.(12),(13), angular dependent 
 phase difference between electron waves, 
 $2 \pi M \omega_c (\theta, \alpha) / \omega_b (\theta)$, 
 is an integer multiple of $2 \pi$ only at 
 commensurate  directions of a magnetic field (4), 
 where $\omega_c(\theta, \alpha) = n \omega_b( \alpha)$, 
with $n$ being an integer.
Therefore, one can conclude that, at arbitrary directions 
of the field (3), the destructive interference effects in 
Eq.(11) result in exponential decay of wave functions 
(9),(10) perpendicular to conducting ${\bf x}$-chains, 
whereas, at commensurate directions (4), the constructive 
interference effects de-localizes electron wave 
functions (9),(10).

To calculate conductivity $\sigma_{\perp} (H, \theta,\phi)$ 
within Fermi liquid approach for non-interacting quasi-particles 
(1), we introduce quasi-classical operator of the velocity 
component $v_{\perp}$ 
in a magnetic field [30]:
\begin{equation}
{\hat v_{\perp}(p_{\perp},x)} = 
- 2 t_{\perp} c^* 
\sin \biggl[ p_{\perp} c^* 
+ \frac{\omega_c(\theta,\phi) x}{v_F} 
- i \ \biggl[ \frac{{\hat \omega_c (\theta , \phi)}}{v_F} \biggl] 
\ \biggl( \frac{d}{d p_y} \biggl) 
\biggl]  \ ,
\end{equation}
and make use of Kubo formalism [7]. 
As a result, we obtain
\begin{equation}
\sigma_{\perp} (H, \theta, \phi) \sim  
\int^{0}_{- \infty} \ d z \
\exp (z) 
\int^{2 \pi}_{0} \frac{d y}{2 \pi} 
\cos \biggl[
\omega_c(\theta,\phi) \tau z 
+ a \biggl( \frac{\cos \phi }{\tan \theta} \biggl)
( f [ \omega_b(\theta) \tau z  + y ] 
- f[y] ) 
\biggl] \ .
\end{equation}

Eq.(15) is the main result of our paper.
Since in Q1D $\rho_{\perp}(H, \theta, \phi) 
= 1 / \sigma_{\perp} (H, \theta, \phi)$,
Eq.(15) solves a problem to determine resistivity
$\rho_{\perp}(H, \theta, \phi)$ for Q1D electron 
spectrum (1) with an arbitrary function 
$f(p_y b^*)$ [37].
Note that the existence of commensurate minima 
(4) in $\rho_{\perp} (H, \theta, \phi)$ directly follows 
from Eq.(15).
Indeed, the discussed above constructive interference 
effects correspond to commensurability of two 
frequencies in Eq.(15), 
$\omega_c(\theta,\phi) = n \omega_b(\theta)$,
where $n$ is an integer. 
Thus, integral (15) increases at commensurate
directions of the field (4) which leads to the
appearance of minima in 
$\rho_{\perp} (H, \theta, \phi)$.

It is possible to make sure that Eq.(15) predicts
non-saturating magnetoresistance at high magnetic 
fields for the case, where wave functions (11)-(13) 
are localized (see Fig.1).
This is in accordance with general theory of
magnetoresistance for open quasi-particle 
orbits [1,7].
Here, we show that suggested in the paper
$1D \rightarrow 2D$ dimensional de-localization
crossovers manifest themselves as saturations 
of magneto-resistance  $\rho_{\perp}(H, \theta, \phi)$ 
at commensurate directions (4) of 
the field (3).
For this purpose, we calculate  
$\rho_{\perp}(H, \theta, \phi) = 
1 / \sigma_{\perp}(H, \theta, \phi)$ as a function 
of $H$ for commensurate angle (4) with 
$N =1$ for the most common case, where 
$f(p_y b^*) = \cos(p_yb^*)$ 
in Eq.(1) [38].
As it is seen from Fig.1, de-localized electrons 
are characterized by unusual for open orbits (1) 
saturated behavior of magnetoresistence,
$ \rho_{\perp}(H, \theta, \phi)  \rightarrow const $
as $H \rightarrow \infty$.   
In Fig.2, we present a comparison of our
theory with the very recent measurements of 
$\rho_{\perp} (H, \theta, \phi)$ performed on 
(TMTSF)$_2$ClO$_4$ [35,38].
Fig. 2 demonstrates that our Eq.(15) is not only 
in qualitative but also in good quantitative agreement 
with the experimental 
results [35,38].

This work was supported in part by National Science 
Foundation, grant number DMR-0308973, the Department 
of Energy, grant number DoE-FG02-02ER63404, and
by the INTAS grants numbers 2001-2212 and 2001-0791.
One of us (AGL) is thankful to E.V. Brusse for 
useful discussions.

\pagebreak

\begin{figure}[h]
\includegraphics[width=6.5in,clip]{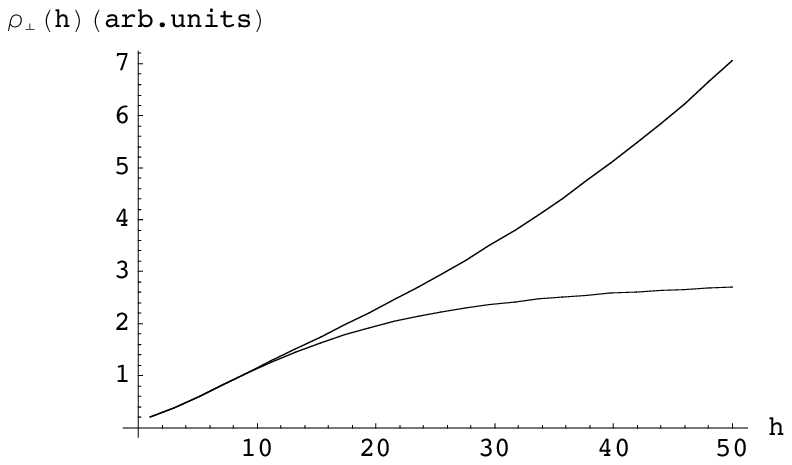}
\caption{ 
Magnetoresistance $\rho_{\perp} (H, \theta, \phi)$
as a function of "dimensionless magnetic field",
$h=\omega_b(H, \theta = \pi/2) \tau$ [see Eq.(7)],
calculated by means of Eq.(15) for Q1D electron 
spectrum (1) with $f(p_y b^*)=\cos(p_y b^*)$,
$t_a/t_b = 8.5$, $b^*=c^*/2.$
Upper curve: non-saturated magnetoresistance
for localized electron wave functions (11)-(13) 
at $\theta =3.5^o$ and 
$\phi = 2.6^o$.
Lower curve: saturated magnetoresistance for
de-localized wave functions (11)-(13)
at $\theta =3.5^o$ and $\phi = 1.75^o$,
corresponding to commensurate direction (4)
of a magnetic field (3) with $n=1$.
}
\label{fig1}
\end{figure}

\begin{figure}[h]
\includegraphics[width=6.5in,clip]{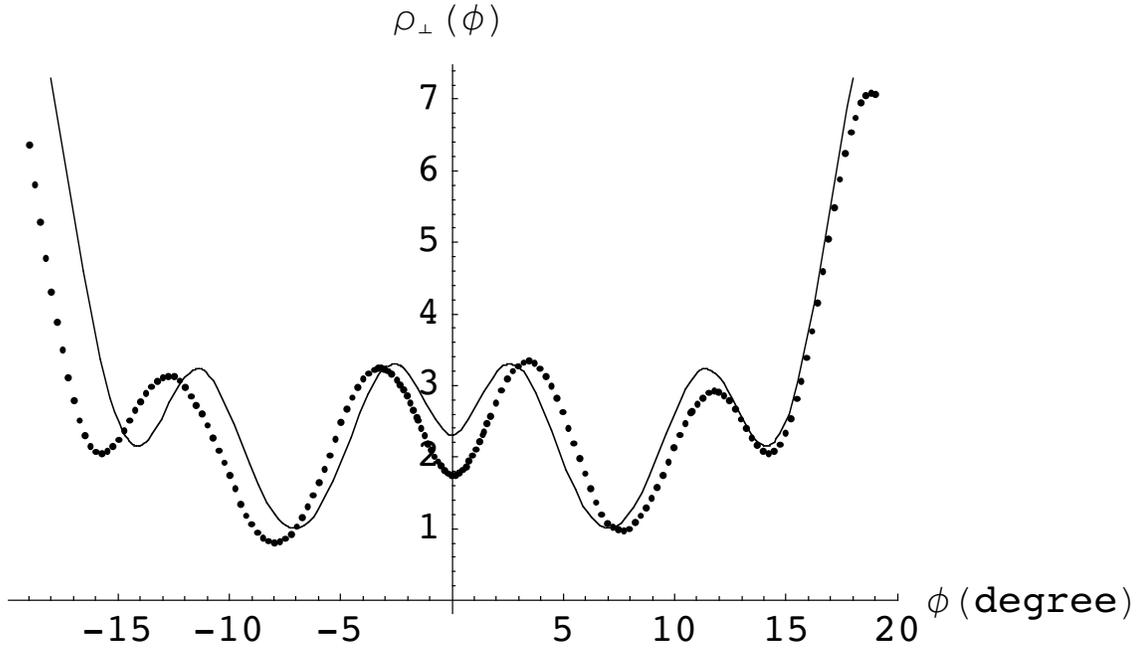}
\caption{ 
Solid curve: magnetoresistance $\rho_{\perp} (H, \theta, \phi)$
as a function of angle $\phi$ calculated at $\theta =7^o$
by means of Eq.(15) for $f(p_y b^*)$ given by Eq.(5) with
$\Delta/2t_b = 0.1$, $t_a/t_b = 10$, $b^*=c^*/2$,
$h=\omega_b(H, \theta = \pi/2) \tau =15$.
Dotes: experimental points [35] obtained on (TMTSF)$_2$ClO$_4$ 
conductor at $\theta =7^o$ and $H = 10 \ T$ [35,38].
}
\label{fig2}
\end{figure}

\pagebreak


\begin{references}
\bibitem{In our} In our opinion, very important from
methodological and historical points of view models [7]
lack of direct physical meanings in Q1D
organic conductors. 
For the discussions, see Ref.[29,28].
\bibitem{Remark-1}  From general equation (15), 
one can obtain its important limiting case for 
$f(p_y b^*)=\cos(p_y b^*)$ considered 
in Refs.[30,31].
\bibitem{Remark-2} For more detailed comparison 
of our Eq.(15) with recent experimental data obtained 
on (TMTSF)$_2$ClO$_4$, 
see Ref.[35].
\end{references}
\end{document}